\documentclass[12pt]{article}
\usepackage{latexsym,graphicx,multirow}
\usepackage{float}
\usepackage{amssymb}
\usepackage{amsmath}
\usepackage{amscd}
\usepackage{amsthm}
\usepackage[left=2cm,top=2.5cm,right=2.5cm,bottom=1.5cm]{geometry}
\usepackage{hyperref}
\usepackage{epstopdf}
\usepackage{cite}
\usepackage{caption}
\usepackage{subcaption}
\usepackage{color}
\usepackage[utf8]{inputenc}
\usepackage[T1]{fontenc}
\begin{document}
	
\begin{center}
\large{\bf{ Exploring the Cosmological Model in $f(R,T^\phi)$ Gravity with Observational Constraints }} \\
\vspace{10mm}
\normalsize{Vinod Kumar Bhardwaj$^1$,  Priyanka Garg$^2$  }\\
\vspace{5mm}
\normalsize{Department of Mathematics, GLA University, Mathura-281 406, Uttar Pradesh, India}\\
\vspace{2mm}
$^1$E-mail:dr.vinodbhardwaj@gmail.com \\
\vspace{2mm}
$^2$E-mail:pri.19aug@gmail.com\\
\vspace{2mm}

\end{center}
	
\begin{abstract}
We have investigated an isotropic and homogeneous cosmological model of the universe in $f(R, T^{\phi})$ gravity, where $T^{\phi}$ is the trace of the energy-momentum tensor and $R$ is the Ricci scalar. We developed and presented exact solutions of field equations of the proposed model by taking the parametrization $q(z) =\alpha + \frac{\beta z}{1+z}$, where $\alpha$ and $\beta$ are arbitrary constants. The best possible values of the model's free parameters are estimated using the latest observational data sets of OHD, BAO, and Pantheon by applying MCMC statistical technique. Some kinematic properties like density parameter $\rho_{\phi}$, pressure $p_{\phi}$, and equation of state parameter $\omega_{\phi}$ are derived.  We have also discussed the behavior of the scalar potential $V(\phi)$ in the $f(R,T^{\phi})$ gravity theory.  The behaviors of scalar fields for quintessence and phantom models are explored. Furthermore, we have discussed the behavior of energy conditions and sound speed in $f(R,T^{\phi})$ cosmology.

\end{abstract}

	
	
\section{Introduction}
Our present universe is in an accelerated expansion era, which has been endorsed by multiple cosmic findings\cite{ref1,ref2,ref3,ref4,ref5,ref6,ref7}. In order to describe this phenomenon within the framework of general relativity, a mysterious energy density known as dark energy (DE) with a large amount of negative pressure has been postulated. Exotic forms of energy can significantly impact the universe's behavior, even though a massive amount of ordinary matter is present, which slows down its expansion. Some researchers have used the reconstruction formula to describe models of dark energy \cite{ref8,ref9}, while others have examined the Type Ia supernova's apparent magnitude-redshift relationship \cite{ref10,ref11}. The basic concepts of dark energy are based on the principles of Einstein's General Relativity, the proper theory of gravity. The cosmological constant has been proposed as a possible substitute for DE due to its repulsive nature, which accounts for the contribution of vacuum energy to the curvature of space-time. However, models based on the cosmological constant have faced fine-tuning problems and cosmic coincidence. \\

In order to explain the accelerating universe, various alternative theories have been proposed without cosmological constant (CC) in the scientific literature. These alternative theories are in the form of modified gravity theories and are based on the modifications of Einstein's general theory of relativity. Modified gravity is an attractive possibility since it provides qualitative solutions to many fundamental issues concerning the late accelerating universe and dark energy. The modified gravity-based DE models can offer viable solutions that are consistent with recent observational data. Some of these which are developed by modifying the Lagrangian and curvature of Einstein’s equation of GR include $f(R)$, $f(T)$, and $f(G)$, and other theories \cite{ref12,ref13,ref14,ref15,ref16,ref17,ref18,ref19,ref20,ref20a,ref20b,ref20c}. Different functions can produce a variety of phenomena in modified theories of gravity. The $f(R)$ theory is considered as the most plausible alternative version of Einstein's theory of general relativity \cite{ref21,ref24,ref25,ref26,ref27}, but it has failed some observational tests, including those in the solar system regime \cite{ref28,ref29}. The $f(R,T)$ theory is another altered form of $f(R)$ theory in which the curvature function is coupled with the function of the energy-momentum tensor. This theory naturally explains the late-time cosmic expansion along with the initial inflation era of universe.
In 2011, Harko et al. \cite{ref30} have proposed a general non-minimal
coupling between matter and geometry in the framework of an effective gravitational Lagrangian consisting of an arbitrary function of the Ricci scalar $R$ and the trace $T$ of the energy–momentum tensor, and introduced $f(R,T)$ gravitational theory. The authors have justified choosing T as an argument for the Lagrangian from exotic imperfect fluids or quantum effects (conformal anomaly). The new matter and time-dependent terms in the gravitational field equations play the role of an effective cosmological constant. A strange behavior of $f(R,T)$ gravity is the non-vanishing covariant derivative of the stress-energy tensor. As a consequence, the equations of motion show the presence of an extra-force acting on a test particle, and consequently the motion is non-geodesic. The authors have applied this theory to analyse the Newtonian limit of the equations of motion and provided a constraint on the magnitude of the extra acceleration by studying the perihelion precession of Mercury. They conclude that the extra acceleration in $f(R,T)$ gravity results not only from a geometrical contribution, but also from the matter content. This extraordinary behavior of $f(R,T)$ gravity has attracted many researchers to explore this theory in different contexts of cosmology and astrophysics \cite{ref30,ref30a,ref31,ref32,ref33,ref34}. The $f(R,T)$ theory well address the issues of objectional viability raised in $f(R)$ theory \cite{ref30,ref30a,ref31}. In context of $f(R,T)$ theory, some exciting studies are mentioned in literature \cite{ref34a,ref34b,ref34c}.
 \\

Besides these theories, the scalar field models are also developed theoretically to depict the inflation era and the present accelerated era of the expanding universe. In these models, the scalar field ($\phi$) is taken as a candidate of the DE. The scalar field produces a negative pressure along with a slow decrease in the potential ($V(\phi)$). Various studies in the literature have been developed to describe the dynamics of the present expanding universe by scalar field cosmologies \cite{ref38,ref39}. The quintessence models are more acceptable scalar field cosmic models which efficiently avoid the well-known issues of fine-tuning and cosmic coincidence and explain the present cosmological scenario of the universe \cite{ref40,ref41}. Initially, the theory of tracking was introduced by Johri \cite{ref42} that suggests a certain path for the evolution of the universe due potential of the tracker. The theory was well justified by observational estimations. Different models of the original quintessence theory have been discussed in literary works. The potential of a scalar field development driven by a non-canonical kinetic term \cite{ref43} and a non-minimal connection between quintessence and dark matter \cite{ref44,ref45} are two examples of these concepts. Numerous fundamental theories also notice the existence of the scalar field phi, which encourages us to research the dynamical characteristics of scalar fields in cosmology. Various cosmological models have been mentioned in literature in the distinct background of scalar field concepts \cite{ref46,ref47,ref48,ref49,ref49a,ref49b}. \\

The $f(R,T^{\phi})$ is a modified version of Einstein's  GR theory \cite{ref30,ref31}. It is based on the modified action principle that includes not only the Ricci scalar curvature $R$ but also a scalar field $T$ that represents the energy-momentum trace. In addition, the theory introduces a scalar field $\phi$ which couples with the matter fields and modifies the gravitational interaction. The theory is characterized by a set of equations that govern the dynamics of gravitational and scalar fields. These field equations are obtained by varying the action with the metric tensor $g^{\mu\nu}$ and scalar fields ($T$ and $\phi$). The resulting equations are highly nonlinear and difficult to solve analytically, but they can be simplified in some cases by adopting a specific kind for the function $f(R,T^{\phi})$. One of the main motivations for considering $f(R,T^{\phi})$ gravity is that they can describe the accelerated expansion universe without the need for dark energy. In addition, the theory can suggest an alternative description for the existence of dark matter and can also lead to modifications of gravity on galactic scales.\\

By considering the parametrization $H(z) = H_0 (1+z)^{1+ \alpha + \beta} exp \left(\frac{- \beta z}{1+z} \right)$, we have analyzed the various parameter by graphical representation. The present study is planned as: We have presents the formulation in $f(R, T^\phi)$ theory and proposed the solution of field equations using parametrization of the deceleration parameter in section 2. The best possible values of model parameters are extracted in section 3. Some dynamics of the proposed model like EoS parameter, scalar field, and potential are discussed in section 4. The sound speed and Energy conditions are also discussed to examine the viability of the derived model in section 4. Finally, the findings for the proposed model are summarized in section $5$.

\section{ Modeling and Solution in $f(R,T^\phi)$ gravity}
For the Einstein's equations, following action is considered in $f(R,T^\phi)$ theory\cite{ref30a,ref31}
\begin{equation}
\label{1}
I=\frac{1}{2}\int(f(R,T^\phi)+2L_\phi)\sqrt{-g}dx^4,
\end{equation}
where $L_\phi=-(\frac{\epsilon}{2}\phi,\mu\phi^,\mu-V(\phi))$, and $8\pi G=1$.\\
By varying action in Eq.(1) w.r.t. $g_{\mu\nu}$, we get the following equation
\begin{equation}
\label{2}
R_{\mu\nu} f_{R}(R,T^\phi)-\frac{1}{2}f(R,T^\phi)g_{\mu\nu}+f_{R}(R,T^\phi)(g_{\mu\nu}\Box-\nabla_\mu\nabla_\nu)=T_{\mu\nu}-(T_{\mu\nu}+\Theta_{\mu\nu})f_T^\phi(R,T^\phi),
\end{equation}
where $\Box= g^{\mu\nu}\nabla_\mu \nabla_\nu$ stands for D'Alembert operator and $\nabla_\mu$ indicates the covariant derivatives w.r.t. $g_{\mu\nu}$ associated with the symmetric Levi-Civita connection.
\begin{equation}
\label{3}
T_{\mu\nu}=-\frac{2}{\sqrt{-g}}\frac{\delta(\sqrt{-g}L_\phi)}{\delta g^{\mu\nu}}=g_{\mu\nu}L_\phi-2\frac{\delta L_\phi}{\delta g^{\mu\nu}},
\end{equation}

\begin{equation}
\label{4}
\Theta_{\mu\nu}=g^{\alpha\beta}\frac{\delta T_{\alpha\beta}}{\delta g^{\mu\nu}}=-2T_{\mu\nu}+g_{\mu\nu}L_\phi-2g^{\alpha\beta}\frac{\delta^2L_\phi}{\delta g_{\mu\nu}\delta g^{\alpha\beta}},
\end{equation}
We have 
\begin{equation}
\label{5}
L_\phi=-\bigg(\frac{\epsilon}{2}\phi,\mu\phi'^\mu-V(\phi)\bigg)=-\frac{1}{2}\epsilon\dot{\phi^2}+V(\phi).
\end{equation}
where dot indicates the time differentiation of the function.\\
From Eq.(4) and Eq.(5), we get 
\begin{equation}
\label{6}
\theta_{\mu\nu}=-2T_{\mu\nu}+g_{\mu\nu}\bigg(-\frac{1}{2}\epsilon\dot{\phi^2}+V(\phi)\bigg).
\end{equation}
From Eqs.(2) and (6), we find 
\begin{equation}
\label{7}
R_{\mu\nu} f_{R}(R,T^\phi)-\frac{1}{2}f(R,T^\phi)g_{\mu\nu}+f_{R}(R,T^\phi) (g_{\mu\nu}\Box-\nabla_\mu\nabla_\nu)=T_{\mu\nu}+(T_{\mu\nu}+g_{\mu\nu}(\frac{1}{2}\epsilon\dot{\phi^2}-V(\phi))) f_{T}^\phi(R,T^\phi).
\end{equation}
Eq.(7) can be rewrite as,
\begin{equation}
\label{8}
Rf_R(R,T^\phi)-2f(R,T^\phi)+3\Box f_R(R,T^\phi)=T^\phi+f_T^\phi(R,T^\phi)(T^\phi+4(\frac{1}{2}\epsilon\dot{\phi^2}-V(\phi))).
\end{equation}
From Eqs. (7) and (8), the Einstein's equation of GR can recasts as 
\begin{equation}
\label{9}
R_{\mu\nu}-\frac{1}{2}Rg_{\mu\nu}=\frac{1}{f_R(R,T^\phi)}(T_{\mu\nu}+T'_{\mu\nu}),
\end{equation}
where
$$T'_{\mu\nu}=f_T^\phi(R,T^\phi)(T_{\mu\nu}+g_{\mu\nu}(\frac{1}{2}\epsilon\dot{\phi^2}-V(\phi)))+\frac{1}{2}(f(R,T^\phi)-Rf_R(R,T^\phi))g_{\mu\nu}+(\nabla_\mu\nabla_\nu-g_{\mu\nu}\Box)f_R(R,T^\phi).$$
Assuming $f(R,T^\phi)$ in the following general form
\begin{equation}
\label{10}
f(R,T^\phi)=-(R+mR^2+nT^\phi).
\end{equation}
For the homogeneous, isotropic $f(R,T^\phi)$ gravity, the metric for spatially flat FLRW model is taken as 
\begin{equation}
\label{11}
ds^2=dt^2-a^2(t)(dr^2+r^2(d\theta^2+\sin^2\theta d\theta^2)).
\end{equation}
For the considered metric, $R_{11}=2\dot{a^2}+a\ddot{a}, R_{22}=r^2(2\dot{a^2}+a\ddot{a})=r^2R_{11}, R_{33}+r^2\sin^2\theta(2\dot{a^2}+a\ddot{a})=r^2\sin^2\theta R_{11}, R_{44}=-3\frac{\ddot{a}}{a}, $, we have 
\begin{equation}
\label{12}
R=-6 \left(\frac{\dot{a^2}}{a^2}+\frac{\ddot{a}}{a} \right).
\end{equation}
For the perfect fluid, the tensor of energy-momentum for a self-interacting scalar potential V($\phi$) and scalar field $\phi$ is 
\begin{equation}
\label{13}
T_{\mu\nu}=\epsilon\phi,\mu\phi,\nu-g_{\mu\nu}[\frac{\epsilon}{2}\phi,\sigma\phi'^\sigma-V(\phi)].
\end{equation}
Using Eq. (13) we get\\
$T_{11}=a^2(\frac{\epsilon}{2}\dot{\phi^2}-V(\phi))$,\\ $T_{22}=a^2r^2(\frac{\epsilon}{2}\dot{\phi^2}-V(\phi))=r^2T_{11}$,\\ $T_{33}=a^2r^2\sin^2\theta(\frac{\epsilon}{2}\dot{\phi^2}-V(\phi))=\sin^2\theta T_{22}$,\\
$ T_{44}=\frac{\epsilon}{2}\dot{\phi^2}+V(\phi) $,\\
 So,
\begin{equation}
\label{14}
T^\phi=-\epsilon\dot{\phi^2}+4V(\phi).
\end{equation}
Using the above equations, the following field equations in $f(R,T^\phi)$ gravity are developed as\cite{ref49c}:
$$2\dot{H}+3H^2-6m(26\dot{H}H^2+2H\dddot{H}+12H\ddot{H}+9\dot{H^2})=(1-n)\frac{\epsilon}{2}\dot\phi^2+(2n-1)V(\phi),$$
\begin{equation}
\label{15}
3H^2-18m(6\dot{H}H^2+2H\ddot{H}-H^2)=(n-1)\frac{\epsilon}{2}\dot{\phi^2}+(2n-1)V(\phi).
\end{equation}
Simplifying above equations, we get the following expressions for $V(\phi)$ and $\dot{\phi^2}$
\begin{equation}
\label{16}
V(\phi)=\frac{6H^2+2\dot{H}-12m(22\dot{H}H^2+9H\ddot{H}+3\dot{H^2}+\dddot{H})}{2(2n-1)},
\end{equation}
\begin{equation}
\label{17}
\dot{\phi}^2=\frac{-2\dot{H}+12m(4\dot{H}H^2+3H.\ddot{H}+6\dot{H^2}+\dddot{H})}{(n-1)\epsilon},
\end{equation}
A strange behavior of $f(R,T)$ gravity is the non-vanishing covariant derivative of the stress-energy tensor. As a consequence, the equations of motion show the presence of an extra-force acting on a test particle, and consequently the motion is non-geodesic.
The study of different phenomena in $f(R,T)$ gravity may also provide some significance signatures and effects which could distinguish and discriminate between various gravitational models. So far, a serious shortcoming of $f(R,T)$ theory is the non-vanishing covariant derivative of the energy–
momentum tensor and, consequently, the standard continuity equation does not hold in this theory, in general. Similarly, the Klein-Gordon equation does not hold if the matter is taken as the scalar field.Since the covariant derivative of the stress-energy tensor does not vanish in general in $f(R,T^\phi)$ gravity, thus the conservation of energy–momentum does not hold.

In FLRW universe, the cosmic pressure and energy density in terms of scalar field $\phi$ are proposed as follows \cite{ref50,ref51} 
\begin{equation}
\label{18}
\rho_\phi=\frac{1}{2}\epsilon\dot{\phi}^2+ V(\phi),
\end{equation}	
\begin{equation}
\label{19}
p_\phi=\frac{1}{2}\epsilon\dot{\phi}^2-V(\phi).
\end{equation}	
For the existing model, the equations of energy density and pressure can be determined as 
\begin{equation}
\label{20}
\rho_\phi=\frac{-n\dot{H}+3(n-1)H^2+6m((-14n+18)\dot{H}H^2)+(-3n+6)H\ddot{H}+(9n-3)\dot{H^2}+n\dddot{H}}{(n-1)(2n-1)},
\end{equation}
\begin{equation}
\label{21}	
p_\phi=\frac{(-3n+2)\dot{H}-3(n-1)H^2+6m((30n-26)\dot{H}H^2+(15n-12)H\ddot{H}+(15n-9)\dot{H}^2+(3n-2)\dddot{H})}{(n-1)(2n-1)}.
\end{equation}	
The equation of state parameter is determined by
\begin{equation}
\label{22}	
\omega\phi= \frac{(-3n+2)\dot{H}-3(n-1)H^2+6m((30n-26)\dot{H}H^2+(15n-12)H\ddot{H}+(15n-9)\dot{H}^2+(3n-2)\dddot{H})}{{-n\dot{H}+3(n-1)H^2+6m((-14n+18)\dot{H}H^2)+(-3n+6)H\ddot{H}+(9n-3)\dot{H^2}+n\dddot{H}}}.
\end{equation}	 
{\large{\bf Parametrization of Deceleration Parameter}}\\
For the considered model, the above physical parameters are the functions of the Hubble parameter. So, to get the explicit solution and examine the dynamics of the universe in $f(R,T^\phi)$ theory, first, we assume the parameterization of the deceleration parameter in the form \cite{ref52}:
\begin{equation} \label{23}
q(z) = \alpha + \frac{\beta z}{1+z}.
\end{equation}
where, $\alpha$ and $\beta$ are constants. $q_{0} = \alpha$ is the current value of deceleration parameter at $z = 0$. For $\alpha = 1/2$ and $\beta = 0$, we have $ q(z) = 1/2$, which describes the dark matter-dominated universe for the derived model. The Hubble parameter in the form of redshift $z$ is defined as
\begin{equation}
\label{24}
H(z) = -\frac{1}{1+z} \frac{dz}{dt}. 
\end{equation}
From Eqs. (23) and (24), the Hubble Parameter is transformed as given below
\begin{equation}
\label{25}
H(z) = H_0 (1+z)^{1+ \alpha + \beta} exp \left(\frac{- \beta z}{1+z} \right).
\end{equation}
Here, $H_0$ denotes the Hubble parameter's current value.

\section{Observational constraints on model parameters}
In this segment, the Markov Chain Monte Carlo (MCMC) process based on Metropolis-Hasting’s procedure has been implemented to determine the best-fit values of model parameters. The $\zeta_{th}$ represents the theoretically predicted value and $\zeta_{ob}$ denotes the corresponding observational value of any observable physical quantity $\zeta$. The $\chi^2$ estimation function is defined as
\begin{equation*}
\label{26}
\chi^{2}_{\zeta}\left(P \right)=\sum_{i=1}^{} {\frac{\left(\zeta_{th}(P)-\zeta_{ob}\right)^2}{\sigma_\zeta^2}}.
\end{equation*}
where $P$ denotes the model parameters and $\sigma_\zeta$ represents the standard deviation in observations of a physical quantity. For the present models, $P = (H_{0}, \alpha,\beta)$ is the parameter vector. By statistically minimizing the estimation function $\chi^2$, the most possible values of model parameters can be determined. We have utilized the observational Hubble data (OHD) consisting of 57 points defined in the range $0.07 \leq z \leq 2.36 $ of redshift \cite{ref53}, observational data from Pantheon compilation, which includes 1048 Type Ia Supernovae (SN Ia) apparent magnitudes in the range of redshift $0.01 \leq z \leq 2.26$ \cite{ref54} and observation dataset of baryon acoustic oscillation (BAO)\cite{ref55,ref56,ref57,ref58,ref59}. The confidence contour plots in two dimensional and marginal plots in one dimension for the model parameters at $1\sigma$ ($68 \%$) and $2\sigma$ ($95\%$) confidence level of the given model are presented in Figure 1. The best-approximated values of parameters for the derived model are shown in Table 1.\\

\begin{table}[H]
	\caption{ The best-fit values of free parameters for different observational datasets }
	\begin{center}
		\begin{tabular}{|c|c|c|c|c|c|}
			\hline\hline 
			\multicolumn{6}{|c|}{ Table of best fit Values for Model }\\
			\hline
			Parameters& $\alpha$ & $\beta$	 & $H_0$  & $z_{t}$ & $q_0$  \\
			\hline
			OHD57 & $-0.571$ & $1.33$  & $68.60$ & $0.752306$ & $-0.571$ \\ 
			\hline
			Pantheon & $-0.676$ & $1.77$  & $71.14$ & $0.617916$ & $-0.676$ \\ 
			\hline
			BAO+OHD57 & $-0.638$ & $1.33$  & $70.00$ & $0.921965$ & $-0.638$ \\ 
			\hline
			BAO+OHD57+Pantheon & $-0.663$ & $1.488$  & $71.17$ & $0.803636$ & $-0.663$ \\
			\hline\hline
		\end{tabular}
	\end{center}
\end{table}
For the combined observational data set, the estimator $\chi^{2}_{total}$ can be expressed as  \begin{equation}
\chi^{2}_{total} = \chi^{2}_{OHD} + \chi^{2}_{Pantheon} + \chi^{2}_{BAO}.
\end{equation}
\begin{figure}[H]
	\centering
	\includegraphics[scale=0.8]{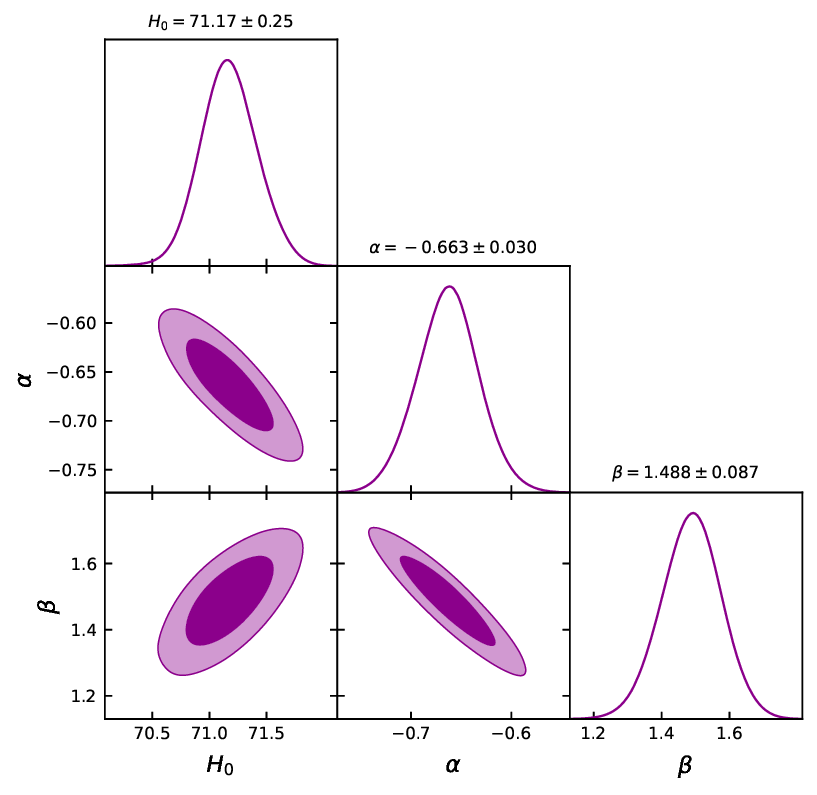} 
	\caption{ Confidence contour plots from OHD+Pantheon+BAO for Model. }
\end{figure}
\section{Features of the Model}
\begin{figure}[H]
	(a)\includegraphics[width=8cm,height=6cm,angle=0]{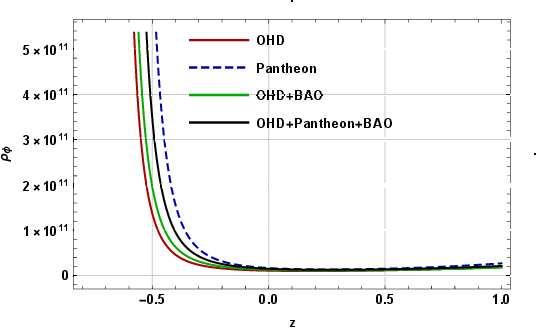}
	~~~(b)\includegraphics[width=8cm,height=6cm,angle=0]{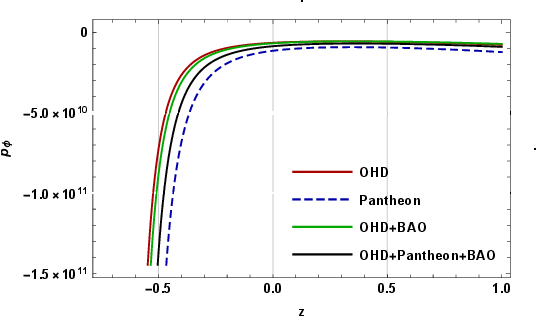}
	~~\caption{(a)  Plots of the energy density vs redshift $z$, (b) Plots of the pressure vs redshift $z$ . }
\end{figure}
The energy density remains positive for different observations data sets as shown in Fig. 2(a) for the mentioned model. The nature of cosmic pressure has been plotted in Figure 2(b) for different observational data sets for the proposed model. The negative behavior of cosmic pressure for the model expresses that the current universe is in the era of accelerated expansion, where the universe was occupied by dark matter in the past. It explains the fact that the space volume increase with the decrease of energy density, and provides an empty space for the present acceleration of the expanding universe.
 
In cosmos studies, the equation of state (EoS) parameter$\omega$ is defined by an association between the dark energy (DE) of the universe and cosmic pressure. The universe attains an accelerated expansion for $\omega < -\frac{1}{3}$. In case of $\omega = -1$, model behaves like $\Lambda$CDM universe. If it lies in the range $-1< \omega \leq 0$ then it lies in the quintessence scenario while for $\omega < -1$, a phantom era is observed. For all best-fitted values, the derived model approved the quintessence era of expanding universe as It has been observed from the EoS parameter. For all best-fitted values of parameters, the $\omega $ lies in the quintessence era, suggesting that the dominant form of the dark energy is believed to be a scalar field, which varies with time. Various cosmological measurements, including observations of the CMB radiation, LSS formation, and Ia supernovae support this scalar field theory\cite{ref3,ref4,ref5,ref6,ref7}.

\begin{figure}[H]
	\centering
	\includegraphics[width=8cm,height=6cm,angle=0]{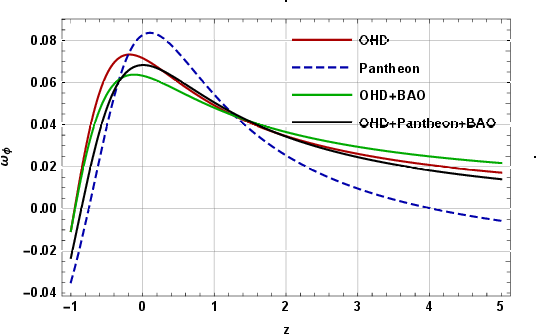}
	\caption{  Plots of EoS parameter vs redshift $z$. }
\end{figure}
\subsection{Scalar potential, Quintessence ($\epsilon = 1$) and phantom ($\epsilon = -1$) Scalar field }
The scalar potential is a concept in physics that describes the energy associated with a scalar field, which has a single value at each space point \cite{ref38,ref39,ref40}. The behavior of a potential and scalar field is dependent on the specific physical system being considered. In general, a positive scalar potential is often associated with stable configurations in physics, as negative energy can lead to unstable or non-physical solutions. However, there are also situations where a negative scalar potential can be physically meaningful, such as in certain models of dark energy or inflation in cosmology. Overall, the specific behavior of a scalar potential depends on the underlying physics and the specific values of the parameters involved. Figure 4 depicts the behaviour of scalar potential. The nature of scalar potential is positive throughout the evolution.
\begin{figure}[H]
	\centering
	\includegraphics[width=8cm,height=6cm,angle=0]{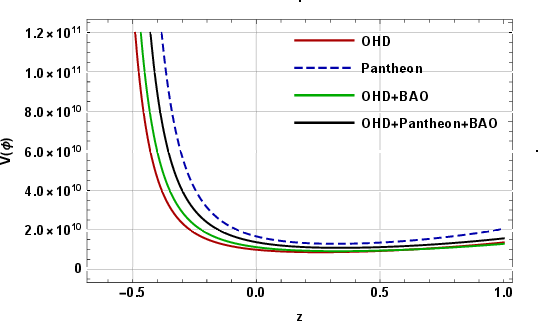}
	\caption{  Plots of  V($\phi$) vs redshift $z$. }
\end{figure}
\begin{figure}[H]
		(a)\includegraphics[width=8cm,height=6cm,angle=0]{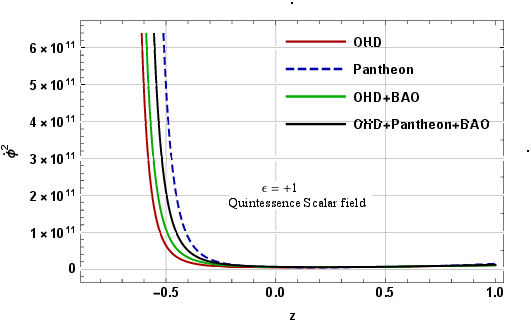}
	~~~(b)\includegraphics[width=8cm,height=6cm,angle=0]{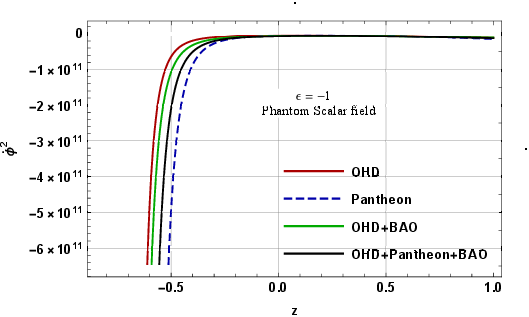}
	~~\caption{(a)  Plots of the V($\phi$) vs redshift, (b) Plots of scalar field vs redshift. }
\end{figure}

A scalar field is a mathematical function that assigns a scalar value (i.e., a single number) to each point in space. Scalar fields are used in many areas of physics, including classical mechanics, electromagnetism, and quantum field theory. Examples of scalar fields include temperature, pressure, and electric potential. \\

From Figure 5(a), we observe that $\dot{\phi}^2$ is positive which means kinetic energy is positive throughout the entire evolution of the cosmos. So, in the quintessence scenario ($\epsilon = +1$), we can expect a massless scalar field $\phi$ with a potential $V(\phi)$, which mimics the gravitational field and effectively depicts the present inflation of cosmos\cite{ref47,ref48,ref49}. In the phantom model ($\epsilon = +1$), a canonical scalar field $\phi$ with negative kinetic energy $\frac{1}{2}\dot{\phi}^2$ possesses a potential that creates a late time accelerated expansion of cosmos. The negative trajectory of kinetic term $\dot{\phi}^2$ for the phantom model depicts the contribution of dark energy for cosmic expansion as shown in figure 5(b) \cite{ref35,ref36,ref37}. 

\subsection{Energy Conditions}

The viability of models is the major issue in the theoretical modeling of the cosmos in the framework of modified theories. In General Relativity, there exist many prevailing energy conditions (ECs) that impose certain conditions to avoid a region whose energy density is negative. These conditions proposed generalized viability to the fact that the energy density of the cosmos always remains non-negative to the entire EMT \cite{ref61a,ref61b}. Many significant space-time, black hole, and wormhole singularity problems can be examined using ECs in a variety of contexts. The well-known Raychaudhari equation \cite{ref61d} can be used to analyze the viability of the numerous ECs that are frequently utilized in GR. The most popular energy condition of GR are termed as Weak energy condition (WEC), Null energy condition (NEC), Dominant energy condition (DEC, and Strong energy condition (SEC). These ECs mathematically can be expressed as: (i)   $\rho +  p \geq 0$ , $\rho \geq 0$ (WEC), (ii) $\rho + p \geq 0$ (NEC), $\rho - p \geq 0$ (DEC), and $\rho + 3 p \geq 0$ (SEC) respectively.\\

Since the energy density and anisotropic pressure for the derived model can be expressed in terms of scalar field $\phi$ means in the form of potential energy $V(\phi)$ and kinetic energy $\frac{\dot{\phi}^2}{2}$. Therefore, ECs of GR in terms of scalar field can be developed as (i) NEC: $\forall \, V(\phi)\geq0$, (ii) WEC $\Leftrightarrow V(\phi)\geq \frac{\dot{\phi}^2}{2}$, (iii) SEC $\Leftrightarrow V(\phi)\geq \dot{\phi}^2$, (iv) DEC $\Leftrightarrow V(\phi)\geq0$. The WEC depicts that the matter-energy density of the cosmos always remains non-negative. The DEC explains that the observed energy density always remains positive and is equal to the measured flux of energy which can never be greater than the light’s speed.  All identified energy and matter forms including dark matter, electromagnetic radiation, and matter particles always satisfy WEC and DEC. The exotic form of energy (DE) that produces a high negative pressure is accountable for the universe’s accelerated expansion and violet the SEC. The violation of the NEC is more exotic and suggests the existence of matter with negative energy density, which is often referred to as "exotic matter." Exotic matter is purely hypothetical and has not been directly observed, but it is often used in theoretical models to explain phenomena such as wormholes and warp drive \cite{ref62,ref63}.
\begin{figure}[H]
	(a)\includegraphics[width=8cm,height=6cm,angle=0]{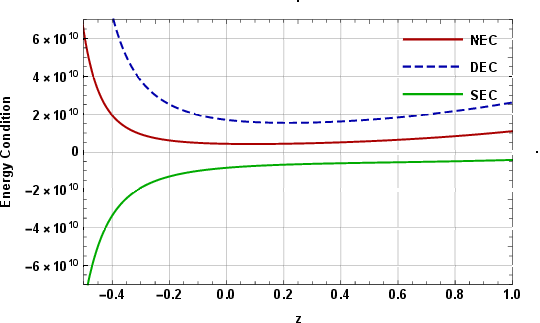}
	~~~~(b)\includegraphics[width=8cm,height=6cm,angle=0]{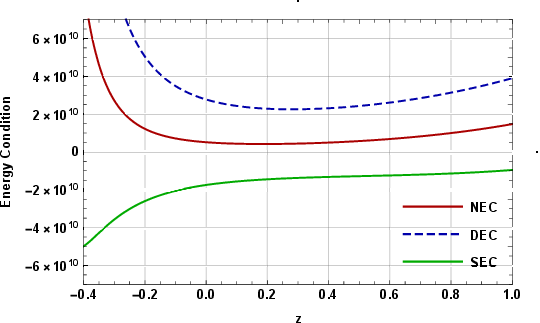}
	(c)\includegraphics[width=8cm,height=6cm,angle=0]{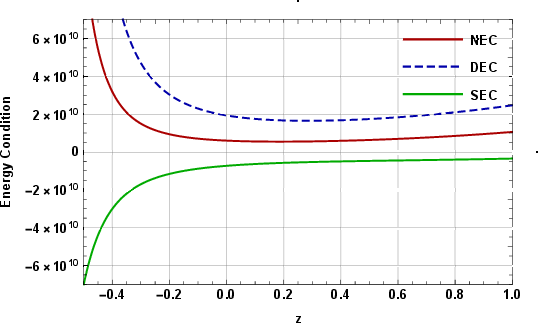}
	~~~~(d)\includegraphics[width=8cm,height=6cm,angle=0]{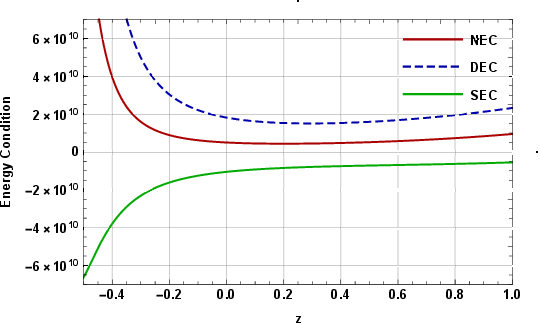}
	~~\caption{ Plots of Energy conditions. }
\end{figure}

For the combined observational data sets of OHD, Pantheon, and BAO, all energy conditions for the derived model are demonstrated in Figure 6. WEC, DEC, and NEC for the specified model are satisfied with the combined data set of observations, as seen in Fig. 6. The violation of SEC for the proposed model led to the universe's accelerated expansion and evidence of exotic matter in the cosmos\cite{ref62a,ref62b}. For $\rho_{\phi}+3 p_{\phi} < 0$, the model lies in inflation era which indicates $V(\phi) > \frac{\dot{\phi}^2}{2}$\cite{ref38,ref39}. 

\subsection{sound speed}

It is required that the velocity of sound $v^{2}_s$ should be less than the velocity of light (c). As we are working in the gravitational units with a unit speed of light, i.e. the velocity of sound exists within the range $0 \leq v^{2}_s = ( \frac{dp}{d\rho}) \leq 1$ with the cosmic time \cite{ref66,ref67,ref68} . The square of sound speed is
\begin{equation}
\label{15}
v^2_{s} = \frac{dp}{d \rho}
\end{equation}
\begin{figure}[H]
 	\centering
 	\includegraphics[width=8cm,height=6cm,angle=0]{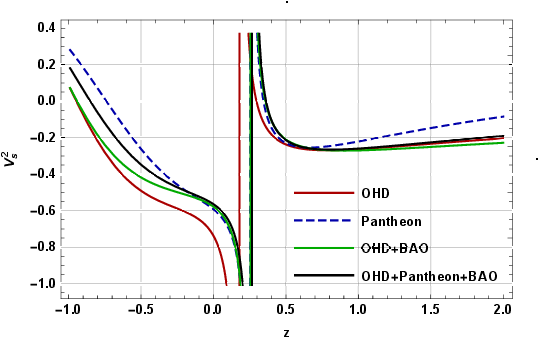}
 	\caption{  Plots of sound speed vs redshift $z$. }
 \end{figure}
For stability of any theoretical model the well-defined range of velocity is $-1 \leq v^{2}_s \leq 1$ for redshift scale. From figure (7), we observed that sound speed lies in well-defined range throughout the evolution of the universe which shows that the model is stable under perturbation since early time except few singularities\cite{ref67,ref68a}.

\section{Concluding remarks}
The paper focuses on the isotropic homogeneous cosmological model with a massless scalar field $\phi$ that follows a potential $V(\phi)$. The paper sets specific values for the arbitrary constants $\alpha$, $\beta$, and $H_0$, and presents graphical representations of various parameters based on these values. The work presented in the paper is summarized as an analysis of this particular cosmological model, including the behavior of its parameters under the specified values of the arbitrary constants. Free parameters of the model are fitted taking OHD, BAO, Pantheon, and their combined data sets using statistical analysis based on the MCMC method. A few kinematic properties like energy conditions, Sound speed are also discussed. The highlights of the model are given below 

\begin{itemize}
	\item
	We have used the observational Hubble data (OHD), observational data from Pantheon compilation, and observation dataset of baryon acoustic oscillation (BAO). Two-dimensional confidence contour plots for the parameters of the derived models are given in Figure-1. The best-estimated values of parameters for the derived models are tabulated in Table 1, for different observational data sets.
	
	\item
	In Figure 2(a), we have depicts the behavior of energy density which is positive during the entire evolution of the cosmos. The nature of cosmic pressure for the proposed model is depicted in Figure 2(b). The nature of pressure is negative for all data sets of the observational findings.
	
	\item
	In Figure 3, the EoS parameter $\omega_{\phi}$ has been plotted for different observational data sets, for the derived model in the $f(R,T^{\phi})$ gravity. Over the entire evolution, it lies in the quintessence region.
	
	\item
	In Figure 4, the scalar potential $V(\phi)$ describes the energy associated with a massless scalar field $(\phi)$. The potential starts with a high value at the initial stage of evolution and gradually decreases, it means that the field is rolling down the potential hill, releasing its energy and approaching a state of minimum energy. This process is known as scalar field inflation and is believed to have played a role in the early universe, driving a rapid expansion known as cosmic inflation. Figures 5(a) and 5(b) show the quintessence and phantom scalar field natures for the derived model.
	
	\item
	In Figure 6, we have estimated the energy conditions for the observational values. We obtain that DEC, and NEC are satisfied while SEC is violated. This indicates that the model under consideration is in the acceleration era of the expanding universe. 
	
	\item
	The nature of the sound speed for the proposed model also points towards stability. The behavior of sound speed for the derived model is plotted in Figure 7, for a combined observational data set.
	
	\item  In the derived model, the accelerated expansion of the universe has been observed due to the violation of conservation	of energy-momentum in $f(R,T^\phi)$ theory.
\end{itemize}

As final concluding remarks, we can say that $f(R,T^\phi)$ gravity
is capable of describing a suitable cosmological model in which a transition from a decelerated to an accelerated phase occurs due to the violation of conservation	of energy-momentum. Hence, the $f(R,T^\phi)$ gravity plays an essential role in the	evolution of the universe. The proposed model turns out viable at present day and n late times observations of $H(z)$,  BAO, and Pantheon data sets. Thus, we have presented a viable a scalar field model in isotropic universe. The proposed model nicely explain the scalar potential driven accelerated expansion of universe.   

\section*{Authors' contributions}
{\bf Vinod Kumar Bhardwaj} : Conceptualization, Ideas, Formulation, Writing  original draft. {\bf Priyanka Garg}: Formal analysis, Methodology, Supervision, Final writing \& editing.
\section*{Data availability} All data generated or analysed during this study are included in this published article.

\section*{Acknowledgement}
The authors wish to express their sincere thanks to esteemed Reviewers and Editors for their constructive comments and suggestions that helped us in the improvement of quality of our work.


\end{document}